\documentclass[ reprint, superscriptaddress, floatfix, amsmath,amssymb, aps ]{revtex4-1}

\usepackage[utf8]{inputenc}
\usepackage{amsmath,amssymb}
\usepackage{wrapfig}
\usepackage{graphicx}
\usepackage{bm}
\usepackage[colorlinks=true,linkcolor=blue,citecolor=blue,urlcolor=blue]{hyperref}
\usepackage{makecell}
\usepackage{float}
\usepackage{caption}
\usepackage{siunitx}

\newcommand{\vect}[1]{\boldsymbol{#1}}
\newcommand{\citeasnoun}[1]{Ref.~\cite{#1}}

\newcommand{\Figref}[1]{Figure~\ref{fig:#1}}
\newcommand{\figref}[1]{Fig.~\ref{fig:#1}}
\renewcommand{\eqref}[1]{Eq.~(\ref{eq:#1})}
\newcommand{\Eqref}[1]{Equation~(\ref{eq:#1})}

\newcommand{\Cref}[1]{Constraint~(\ref{c:#1})}

\newcommand{\eqreftwo}[2]{Eqs.~(\ref{eq:#1},\ref{eq:#2})}

\newcommand*{\Ev}{\vect{E}}
\newcommand*{\xv}{\vect{x}}

\newcommand*{\diag}[1]{\textrm{diag}\left(#1\right)}
\newcommand*{\SM}{SM}
\newcommand*{\epso}{\varepsilon_{\rm origin}}

\begin{document}
\title{Minimum Dielectric-Resonator Mode Volumes}
\author{Qingqing Zhao}
\affiliation{Department of Applied Physics and Energy Sciences Institute, Yale University, New Haven, Connecticut 06511, USA}
\affiliation{Department of Physics, University of Hong Kong, Hong Kong, China}
\author{Lang Zhang}
\affiliation{Department of Applied Physics and Energy Sciences Institute, Yale University, New Haven, Connecticut 06511, USA}
\author{Owen D. Miller}
\affiliation{Department of Applied Physics and Energy Sciences Institute, Yale University, New Haven, Connecticut 06511, USA}

\date{\today}

\begin{abstract}
    We show that global lower bounds to the mode volume of a dielectric resonator can be computed via Lagrangian duality. State-of-the-art designs rely on sharp tips, but such structures appear to be highly sub-optimal at nanometer-scale feature sizes, and we demonstrate that computational inverse design offers orders-of-magnitude possible improvements. Our bound can be applied for geometries that are simultaneously resonant at multiple frequencies, for high-efficiency nonlinear-optics applications, and we identify the unavoidable penalties that must accompany such multiresonant structures.
\end{abstract}

\maketitle


Resonators that confine electromagnetic waves to highly subwavelength regions of space~\cite{Eli1998,Lipson05,Gondarenko2008,Liang2013,Hu2016,Choi2017,Yang2017,Hu2018,Wang2018,David19} are useful for applications ranging from novel light sources~\cite{Altug2006,Matsuo2010} and high-efficiency nonlinear optics~\cite{nonlinear,Nozaki2010,Pant2011} to cavity QED~\cite{QED,Englund2007,Tiecke2014}, yet the \emph{maximal} confinement of a mode is not known: perfectly sharp tips support field singularities with zero mode volume, but fabrication constraints prevent perfect sharpness. In this Letter, we identify global lower bounds to the mode volumes of high-$Q$ dielectric resonators, using Lagrangian duality~\cite{Boyd2004,Angeris2019} and convex optimization to reveal bounds that depend only on the material refractive index and minimum achievable feature size. For two-dimensional subwavelength confinement, as is typical in lithographically defined structures, we find that the bounds scale quadratically with minimum feature size. Surprisingly, state-of-the-art designs~\cite{Hu2016,Choi2017,Hu2018} based on bowtie-antenna-like sharp tips exhibit only linear scaling, falling short of the bounds by 20X at $\lambda/50$ minimum feature sizes, and more than 100X at single-nanometer feature sizes and telecommunications wavelengths. We show that more complex structures discovered by ``inverse design''~\cite{Jensen2011,miller2012photonic,Bendsoe2013,Molesky2018} show superior scaling and appear capable of orders-of-magnitude improvement over sharp-tip-based designs. For scalar waves (such as TE electromagnetic modes, acoustic waves, or single-particle quantum wavefunctions), which cannot utilize the discontinuities arising from the vector-Maxwell boundary conditions, we find bounds that are nonzero for arbitrarily small features yet still significantly below the half-wavelength ``limit.'' We also show that this computational technique can discover bounds for dual- and multi-frequency-resonant structures, an important class of structures for nonlinear frequency-conversion applications~\cite{NonlinearConversion,Bi2012,Buckley2014,Lin2016,Lin2017,Sitawarin2018} and a regime where analytic-continuation-based bound techniques~\cite{Sohl2007,Hashemi2012,Shim2019} offer no help. Our framework applies to all linear wave resonators and demonstrates the power of computational-optimization techniques for identifying global bounds in high-dimensional design spaces.


Electromagnetic resonators with highly subwavelength mode volumes $V$ exhibit scattering responses proportional to $1/V$ (or higher powers thereof) when excited by near- or far-field sources, as in the Purcell effect~\cite{Purcell1946,Novotny2012}. In a nonmagnetic dielectric medium with permittivity $\varepsilon$, the mode volume $V$ of a high-$Q$ modal field $\Ev$ is the ratio of the total field energy to the energy at a maximum-intensity point $\xv_0$, $\int \varepsilon |\Ev|^2 / \varepsilon(\xv_0) |\Ev(\xv_0)|^2$~(\citeasnoun{Novotny2012}). Electromagnetic-field discontinuities across interfaces enable highly subwavelength mode volumes~\cite{Lipson05,Gondarenko2008} that can be designed by computational optimization~\cite{Liang2013,Hu2016,Wang2018,Hu2018} or quasistatic self-similarity~\cite{Choi2017}.

In parallel there has been significant effort towards discovering analytical bounds, or fundamental limits, across a wide variety of electromagnetic response functions~\cite{gordon_1963, purcell_1969,miller2000communicating,miller2007fundamental,Sohl2007,kwon2009optimal, Raman2013, liberal2014least, liberal2014upper,miller_fundamental_2016,miller2017limits,yang2017low, hugonin_fundamental_2015,miller2015shape, rahimzadegan2017fundamental,miller2017universal,Milton2017,liu2018optimal,yang2018maximal,michon2019limits, nordebo2019optimal, ivanenko2019optical,dias_fundamental_2019,Shim2019,Kuang2020}. There are bounds on local densities of states~\cite{miller_fundamental_2016,Shim2019}, which for a single resonator is proportional to $Q/V$, but for lossless materials these bounds take arbitrarily large values or diverge. Recently, it has been recognized that \emph{computational} bounds are also possible~\cite{Angeris2019,Gustafsson2019,Kuang2020,Molesky2020,Trivedi2020}, via Lagrangian duality~\cite{Boyd2004}. Particularly relevant is \citeasnoun{Angeris2019}, which develops a duality-based approach to bounding least-squares error between any designable field and an ideal target field. However, use of a target field prohibits bounds on a response function itself, as the squared-error objective is only an error metric; moreover, it is rare for the ideal target field itself to even be known. The minimum-mode-volume problem, through suitable transformations described below, has a target-field-free Lagrangian dual formulation.

\begin{figure} 
    \centering
    \includegraphics[width=0.6\linewidth]{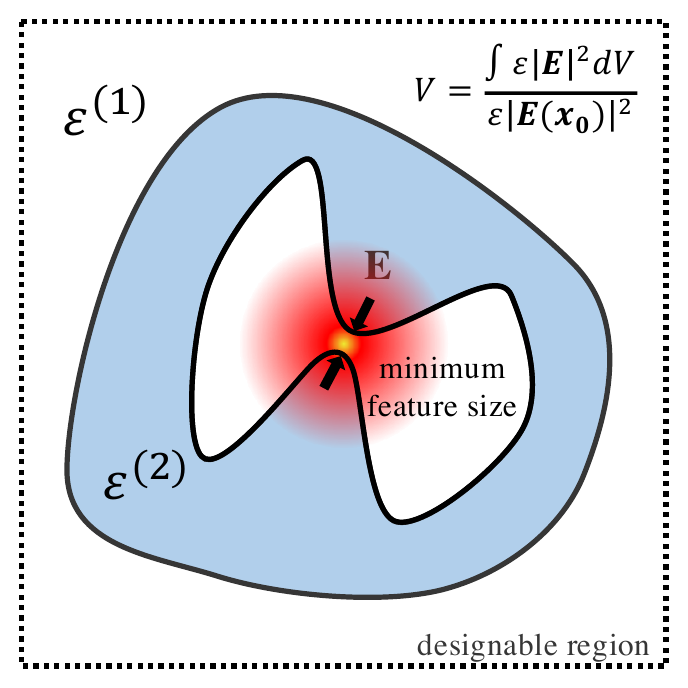}
    \caption{Schematic of a patterned dielectric resonator with minimum feature size $d$ supporting a resonance with small mode volume $V$. We use Lagrangian duality to find lower bounds on the mode volume, over all possible geometrical configurations.}
    \label{fig:schematic}
\end{figure}

\emph{Dual formulation}--The smallest mode volume of a dielectric resonator with resonant frequency $\omega$ is the solution of a minimization problem over all allowed permittivity distributions $\varepsilon(\xv)$ and electric fields $\Ev(\xv)$:
\begin{equation}
\begin{aligned}
& \underset{\varepsilon,\Ev}{\text{minimize}}
& & V = \frac{\int \varepsilon(\xv) |\Ev(\xv)|^2\,{\rm d}\xv}{\varepsilon(\xv_0) \left| \Ev(\xv_0) \right|^2} \\
& \text{subject to}
& & \nabla \times \nabla \times \Ev = \omega^2 \varepsilon \Ev.
\end{aligned}
\label{eq:minproblem}
\end{equation}
We depict this problem schematically in \figref{schematic}. Without any restrictions on the permittivity distribution, the solution is trivially 0, as perfectly sharp tips which enclose dielectric (or metallic) materials at angles less than \ang{180} support integrable field singularities~\cite{dielectricWedge,Choi2017}. Thus we only consider permittivity distributions with a minimum feature size $d$, and we do not allow any edges to approach within $d/2$ of $\xv_0$. 

To enable a dual formulation of \eqref{minproblem}, we transform the problem in three ways. First, we treat the total field intensity $\int \varepsilon \left|\Ev\right|^2$ as the minimization metric, as it is biconvex in $\varepsilon$ and $\Ev$, and we constrain the field intensity at the origin, $\varepsilon(\xv_0) \left| \Ev(\xv_0) \right|^2$, to be 1. Second, we introduce a perfectly matched layer (PML) to restrict the problem to a finite region, and we define a weight function $W(\xv)$ that is unity everywhere except the PML, where it takes a small but nonzero value (important for invertibility below). Third, we ``lift'' the problem to a higher-dimensional setting to linearize the newly introduced field-intensity constraint: instead of fixing $\varepsilon(\xv_0) |\Ev|^2 = 1$, we fix $\sqrt{\epso} \hat{\vect{p}} \cdot \Ev(\xv_0) = 1$, where $\epso$ is a binary value taking one of two possible values and $\vect{p}$ is a polarization vector that we optimize over. Finally, we assume any standard discretization scheme to reduce the problem to a finite-dimensional one~\cite{Jin2011}, we separate the real and imaginary parts of all variables and treat them as independent degrees of freedom, and we represent vectors in lowercase and matrices in uppercase. These operations, detailed in the {\SM}, transform \eqref{minproblem} to the problem
\begin{equation}
\begin{aligned}
& \underset{\varepsilon, \epso, e, v}{\text{minimize}}
& & e^T W^T \diag{\varepsilon} W e \\
& \text{subject to}
& & A e  = \diag{\varepsilon} e \\
& & & \sqrt{\epso} v^T e = 1,
\end{aligned}
\label{eq:genminproblem}
\end{equation}
where $A$ is the discrete representation of a frequency-normalized curl-curl operator ($(1/\omega^2) \nabla \times \nabla \times$), $v$ is the discrete representation of a delta function at $\xv_0$ with amplitude $\hat{\vect{p}}$, $e$ is the discretized electric-field vector, and ``diag'' denotes the matrix with its argument on the diagonal and all zeros otherwise. The permittivity is constrained to lie between a background value, $\varepsilon^{(1)}$, and the resonator-material value $\varepsilon^{(2)}$. A minimum feature size $d$ can be enforced by partitioning the geometry into a disjoint set of size-$d$ elements and requiring constant permittivity across each element. We further simplify \eqref{genminproblem} by concatenating the two linear constraints (cf. {\SM}).

Every ``primal'' minimization problem has a dual function that lies entirely below the minimal value of the primal problem, so that its maximum serves as a potentially optimal lower bound for the primal problem~\cite{Boyd2004}. A dual function is always concave, independent of the convexity of the primal problem; thus its maximum is the solution of a convex-optimization problem and can be solved reliably and efficiently~\cite{Boyd2004}. However, the dual function is itself the solution of an optimization problem, and in many scenarios it is as difficult to solve as the primal problem. Here we show that our formulation of the mode-volume problem via \eqref{genminproblem} leads to a semi-analytical form of the dual function that is amenable to rapid maximization.

The variables $\epso$ and $v$ occupy low-dimensional spaces ($\epso$ is binary and $v$ depends only on two angles) and can be treated as fixed parameters within an ``inner-loop'' optimization over $\varepsilon$, and then optimized themselves in an ``outer-loop'' grid search. In the inner-loop minimization, the Lagrangian function is given by:
\begin{align}
    L(\varepsilon,e,\nu) = e^T W^T \varepsilon W e + \nu^T \left[\left(A-\varepsilon\right) e - b\right] + I(\varepsilon),
\end{align}
where $I$ is an indicator function that is zero for valid permittivity distributions and $+\infty$ otherwise. To find the dual function, the next step is to minimize over the primal variables $E$ and $\varepsilon$. We use a modified version of the derivation presented in \citeasnoun{Angeris2019}, as detailed in the {\SM}. After introduction of an auxiliary vector variable $t$, the dual problem is 
\begin{equation}
    \begin{split}
        \mathop{\text{maximize}}_{\nu,t}\quad &-\frac{1}{4}\vect{1}^T t-\nu^T b\\
        \text{subject to}\quad &t_i\geq\sum_{j\in S_i}\frac{((A^T\nu)_j-\varepsilon_i^{(k)}\nu_j)^2}{{W_{j,j}^2\varepsilon_i^{(k)}}}, \\ &\quad \forall i\in[m], \quad k\in\{1,2\}\\
    \end{split}
    \label{eq:dual}
\end{equation}
where $j$, $i$, and $k$ are indices for the individual pixels, the fabrication blocks, and the possible permittivity values, respectively, and $m$ is the total number of fabrication blocks. A technical point in our specification of \eqreftwo{minproblem}{genminproblem} is that we specify $\omega$ to be the real-valued frequency of interest, with no imaginary part, as our numerical experiments show that bounds converge to minimal values in this limit (\SM). \Eqref{dual} is a convex quadratically constrained quadratic program; we use the modeling language CVX~\cite{cvx} to rewrite it as a second-order cone program (SOCP) and solve it with the Gurobi solver~\cite{gurobi}.
\begin{figure}
    \centering
    \includegraphics[width=0.8\linewidth]{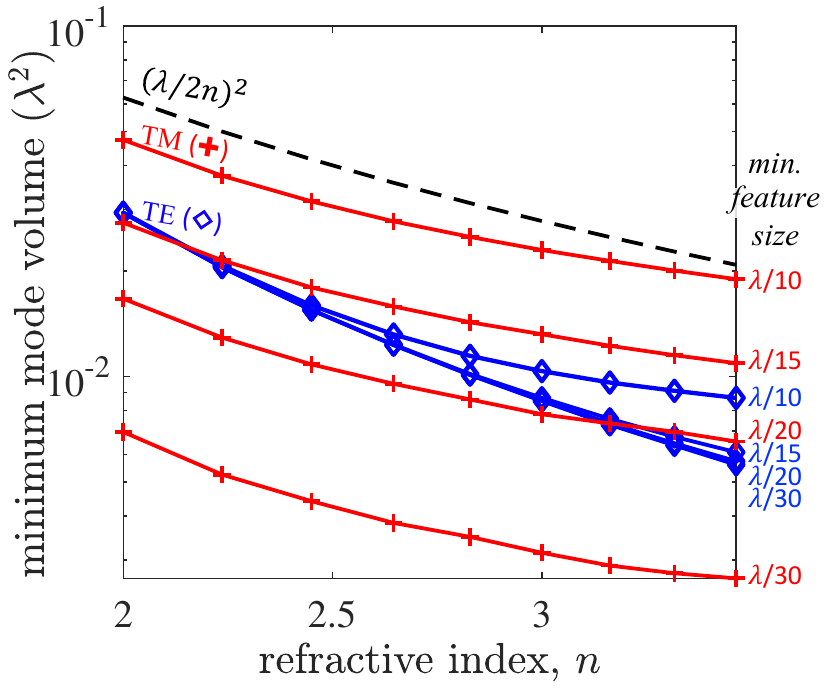}
    \caption{Minimum mode volumes for 2D TE (scalar) and TM (vector) waves, for minimal feature sizes from $\lambda / 10$ to $\lambda / 30$. All of the bounds lie below the ``diffraction limit'' $(\lambda/2n)^2$ (dashed black line).}
    \label{fig:bddVSn}
\end{figure}

\emph{Mode-Volume Bounds}---\Figref{bddVSn} depicts mode-volume bounds computed from \eqref{dual} as a function of refractive index, considering two-dimensional resonators that serve as prototypes for lithographically defined physical structures. A design region of size $3\lambda \times 3\lambda$ for wavelength $\lambda$ is considered; the mode-volume bounds rapidly converge for diameters beyond roughly $1\lambda$ (cf. {\SM}), as distant scatterers can modify quality factor significantly but not field intensity at the origin. The bounds decrease with refractive index, as expected, but the effect of varying the minimum feature size is highly polarization-dependent. For TE modes, the field is continuous across the dielectric boundaries, and sharp tips do \emph{not} exhibit diverging fields~\cite{dielectricWedge}, so the mode volume remains nonzero even for arbitrarily small feature sizes. By contrast, for TM modes the electric field can be highly discontinuous across boundaries and divergent at sharp tips, which is the underlying mechanism exploited for previous deep-subwavelength designs~\cite{Lipson05,Choi2017,Liang2013}. Also included in \figref{bddVSn} is the ``diffraction limit,'' $(\lambda/2n)^2$, which can be significantly improved upon even with relatively large minimum feature sizes. The bounds of \figref{bddVSn} are global bounds and cannot be surpassed through any kind of structural design.
\begin{figure*}[hbt]
    \includegraphics[width=0.9\textwidth]{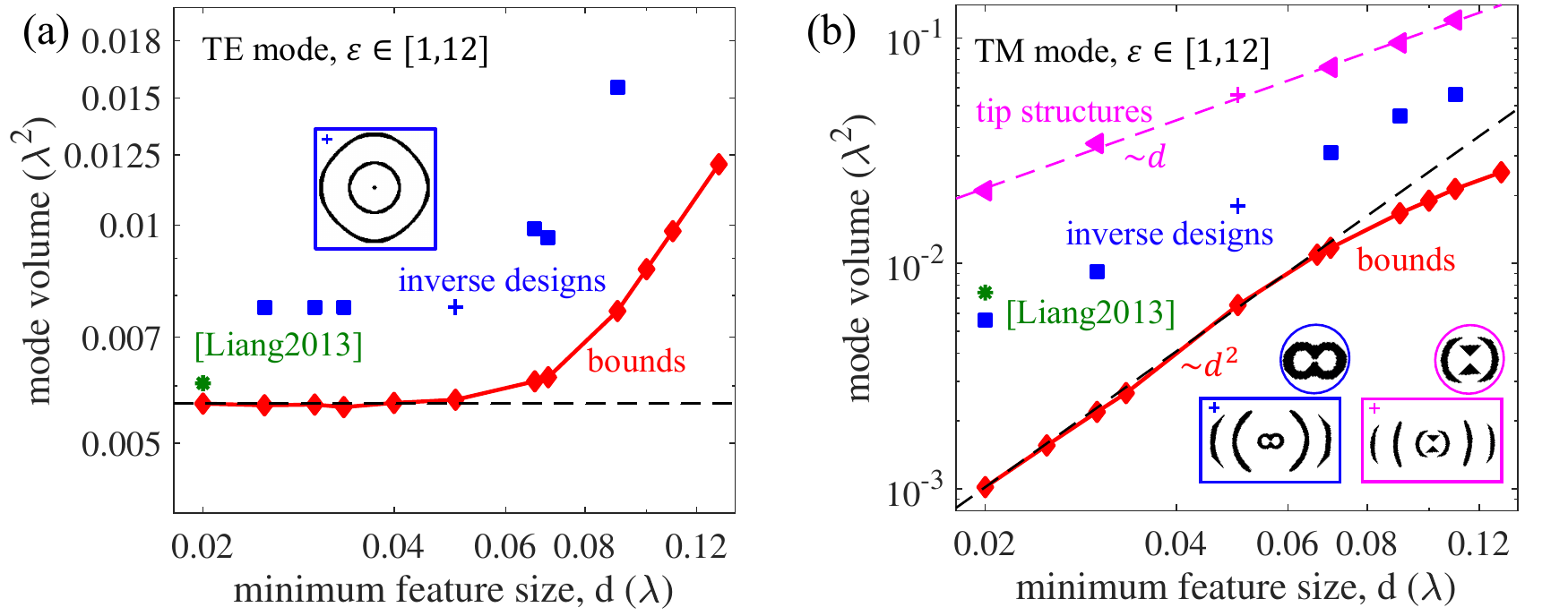}
    \centering
    \caption{Comparison of mode-volume bounds (red) and inverse-designed resonators (blue, green) as a function of minimum feature size, for (a) TE and (b) TM modes. At very small feature sizes the bounds converge to asymptotic values that are constant for TE modes and scale as $d^2$ for TM modes (black dashed lines). For TM modes, we also include optimized bowtie-antenna-like ``tip'' structures. The tip structures (pink) show a linear scaling that diverges from the bounds, while the inverse designs track them more closely.}
    \label{fig:bddVSfab}
\end{figure*}

A natural question is how ``tight'' the bounds are---can designed structures come close to the bounds? \Figref{bddVSfab} depicts the mode-volume bounds as a function of minimum feature size for materials $\varepsilon^{(2)} = 12$ and $\varepsilon^{(1)}= 1$ (typical of silicon and air) for both TE and TM polarizations. Alongside the TM bounds we include a series of data points that arise from using structures based on sharp tips (pink markers), which have been the basis of many state-of-the-art designs~\cite{Choi2017,Liang2013}, which perform well at larger feature sizes but show a large gap (20X) from the bounds at smaller feature sizes ($\lambda/50$). To find superior designs, we use ``inverse design''~\cite{Jensen2011,miller2012photonic,Bendsoe2013,Molesky2018} to discover optimal structures. We employ ``topology optimization,'' wherein the refractive index at any pixel is a variable parameter, with a penalty function that ultimately enforces binary designs. (More computational details are included in the {\SM}.) The results of these optimizations, alongside representative designs, are included as blue markers and a blue inset in \figref{bddVSfab}. For both polarizations, and across many features sizes, the inverse-designed structures approach within a factor of 2--5 of the bounds. \citeasnoun{Liang2013} used a sophisticated contour-integration-based inverse-design approach to discover high-quality-factor, small-mode-volume structures, and their designs (green markers) also come quite close to the bounds (within $7\%$ for TE polarization).

In \figref{bddVSfab}, the TM-mode bounds scale quadratically with feature size $d$ (black dashed line), whereas tip-based designs only exhibit linear scaling (pink dashed line). We can extend this analysis to 3D structures to compare state-of-the-art designs~\cite{Hu2018,Choi2017} to our bounds. Solving \eqref{dual} can require significant computation times (7 hours on a 20-core machine even for a single 2D TM bound), which for 3D bounds will require new software implementations that are out of the scope of this paper (CVX is not optimized for large-scale problems). Yet we can apply a ``2.5D''~\cite{Lu2011} analysis to make predictions. In prototypical on-chip implementations, waveguide modes occupy approximately $\lambda/2n$ mode thickness in the third dimension, with all subwavelength confinement arising from the two in-plane directions. Extrapolating the linear scaling of \figref{bddVSfab} to the feature sizes of Refs.~\cite{Hu2018,Choi2017}, with $\lambda/2n$ confinement in the third dimension, we predict mode volumes within a factor of 2 of their simulated values, cf. Table~\ref{3Dbound}. According to the same table, our bounds suggest the possibility for two orders of magnitude improvement.

\begin{table}[h]
\renewcommand{\arraystretch}{1.3}
\begin{tabular}{|c|c|c|c|}
\hline
    Gap & 3D design & 2.5D tip estimate & 2.5D bound \\
\hline
    \SI{1}{nm} & $7\times 10^{-5}\lambda^3$~\cite{Choi2017} & $10\times 10^{-5}\lambda^3$ & $2\times 10^{-7}\lambda^3$ \\
\hline
    \SI{4}{nm} & $2\times 10^{-4}\lambda^3$~\cite{Hu2018} & $4\times 10^{-4}\lambda^3$ & $2\times 10^{-6}\lambda^3$ \\
\hline
\end{tabular}
\caption{State-of-the-art theoretical designs (``3D designs'') have achieved mode volumes near or below $10^{-4}\lambda^3$ at sub-\SI{5}{nm} gap distances for wavelengths near \SI{1550}{nm}. These are within a factor of 2 of ``2.5D'' extrapolations from the linear scaling of \figref{bddVSfab}(b) for tip-based structures. Yet 2.5D extrapolations of the bounds (``2.5D bound'') suggest the possibility for two orders of magnitude improvement.}
\label{3Dbound}
\end{table}

\begin{figure}
    \centering
    \includegraphics[width=0.8\linewidth]{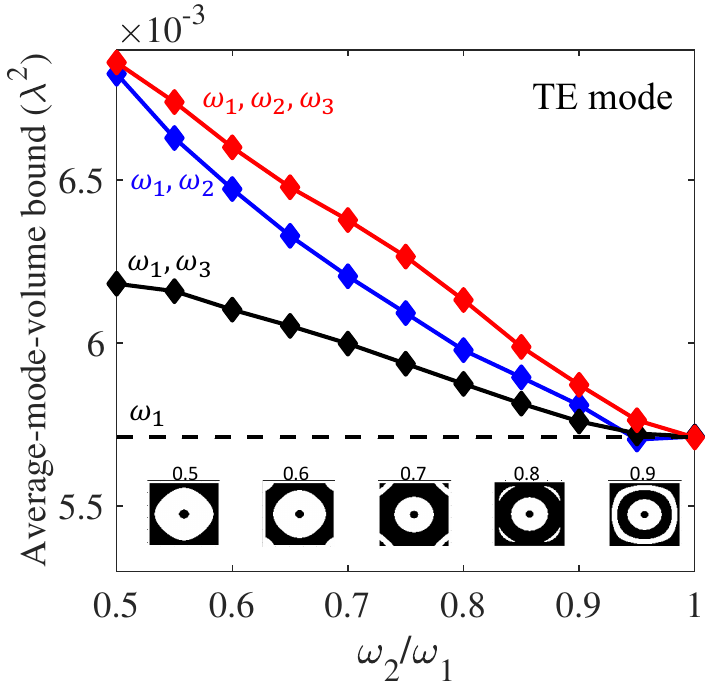}
    \captionof{figure}{Lower bounds for structures that support multiple resonances, e.g., for high-efficiency nonlinear optics,  at frequencies $\omega_1$, $\omega_2$, and/or $\omega_3$, with $2\omega_1 = \omega_2 + \omega_3$. Relative to the single-resonance bounds (black dashed line), the dual-resonance (blue, black) and triple-resonance (red) structures have larger bounds, exhibiting the penalty associated with requiring multiple resonances. Inset: Structures extracted from the dual-program optimization, labeled by the value of $\omega_2$ for which they are computed.}
    \label{fig:multimode}
\end{figure}

A unique feature of this computational bound approach is its capability to identify bounds for \emph{multiresonant} structures that can be particularly important for enhancing nonlinear-optic effects~\cite{NonlinearConversion,Bi2012,Buckley2014,Lin2016,Lin2017,Sitawarin2018}. The condition of requiring resonances at a set of multiple frequencies, $\{\omega_1, \omega_2, \ldots\}$, simply represents additional biconvex constraints in \eqref{genminproblem}, resulting in a larger version of the dual problem of \eqref{dual}. As an example, we consider nonlinear-optical processes involving three frequencies $\omega_1$, $\omega_2$, and $\omega_3$ that satisfy the condition $2\omega_1 = \omega_2 + \omega_3$, and in \figref{multimode} we vary the value of $\omega_2$ relative to $\omega_1$. The constant dashed line in \figref{multimode} is the single-frequency bound for a resonator supporting a resonance at $\omega_1$. The black and blue lines represent the computed bounds on average mode volume for dual-frequency cavities (cf. {\SM} for details), at frequencies $\omega_1$ and $\omega_2$ (blue) or $\omega_1$ and $\omega_3$ (black), which represent cavities that could be used for enhancement of second-order $\chi^{(2)}$ nonlinear processes. The bounds for these dual-resonant cavities are larger than those of the single-frequency cavity, representing a penalty for requiring simultaneous resonances at two frequencies. Finally, the largest mode-volume bounds, shown in red, are for cavities resonant at each of \emph{three} frequencies $\omega_1$, $\omega_2$, and $\omega_3$, which can enhance third-order, $\chi^{(3)}$ nonlinear response. For $\omega_2 = 0.5\omega_1$, the average-mode-volume bound is about 20\% larger than the single-frequency bound, illustrating the penalty associated with multiresonant structures, and the unique ability of this approach to identify such penalties.

\emph{Conclusion}--This Letter presents a method for computing global lower bounds to the mode volume of a dielectric resonator, naturally accounting for fabrication and multi-frequency constraints that are particularly difficult to incorporate into analytical electromagnetic-response bounds. One can imagine the potential utility of such a technique beyond mode volume, for applications ranging from metasurfaces~\cite{yu2014flat,aieta2015multiwavelength,chung2019tunable} to quantum nanophotonics~\cite{QED,QPhotonics}. A key hurdle in extending this approach to such applications will be retaining the positive-definite quadratic-form properties of the objective function, to maintain the feasibility of finding the Lagrangian dual. Alternatively, it is possible that one may find an analytical dual function for more general objective forms. \citeasnoun{Trivedi2020} makes progress on identifying dual functions for particular scattering problems, albeit for small resonators. Simultaneously, there has been significant progress in identifying convex and nonconvex quadratic constraints on the polarization currents excited in any scattering problem~\cite{Gustafsson2019,Kuang2020,Molesky2020} and utilizing them for bounds. A convergence of these approaches may lead to a continuum of analytical, semi-analytical, and computational bounds, incorporating varying levels of information, for any electromagnetic response function.


\bibliography{paper}
 
\end{document}


\title{Supplementary Material: Minimum Dielectric-Resonator Mode Volumes}
\author{Qingqing Zhao}
\affiliation{Department of Applied Physics and Energy Sciences Institute, Yale University, New Haven, Connecticut 06511, USA}
\affiliation{Department of Physics, University of Hong Kong, Hong Kong, China}
\author{Lang Zhang}
\affiliation{Department of Applied Physics and Energy Sciences Institute, Yale University, New Haven, Connecticut 06511, USA}
\author{Owen D. Miller}
\affiliation{Department of Applied Physics and Energy Sciences Institute, Yale University, New Haven, Connecticut 06511, USA}

\date{\today}

\maketitle

\tableofcontents

\section{The optimization problem and its dual problem}
The optimization problem is to maximize the mode volume $ \frac{\int \varepsilon |\Ev|^2}{\varepsilon(\xv_0) \left| \Ev(\xv_0) \right|^2}$ under the Maxwell constraint, where variable $\varepsilon(\xv)$ and $\Ev(\xv)$ are the permittivity distribution and modal electric field respectively. The parameter $\xv_0$ is the point with maximal energy density and is set to be at the origin through out our derivation. The minimal mode volume problem can thus be written as \eqref{minproblem}. 
\begin{equation}
\begin{aligned}
& \underset{\varepsilon, \Ev}{\text{minimize}}
& & V = \frac{\int \varepsilon |\Ev|^2}{\varepsilon(\xv_0) \left| \Ev(\xv_0) \right|^2} \\
& \text{subject to}
& & \nabla \times \nabla \times \Ev = \omega^2 \varepsilon \Ev.
\end{aligned}
\label{eq:minproblem}
\end{equation}
The parameter $\omega$ is a real number representing the resonate frequency of the mode that we optimize. The permittivity $\varepsilon$ is constrained to lie between a background value, $\varepsilon^{(1)}$, and the resonator-material value $\varepsilon^{(2)}$. To find the global bound for the dielectric-resonator mode volume, we first perform several transformations on \eqref{minproblem}. We then demonstrate how to find the dual problem of the transformed problem using procedures similar to those in \citeasnoun{Angeris2019}.

A typical formulation of the Maxwell eigenproblem of \eqref{minproblem} specifies an integral normalization, e.g., $\int \varepsilon |\Ev|^2 = 1$ (\citeasnoun{PhotonicCrystal}). As implemented in \eqref{minproblem}, such a constraint would fix the numerator of the objective and leave the origin intensity in the denominator as the objective function. Yet such objective functions are hard to work with when searching for analytical representations of the dual function and we made no progress with that form of the problem. But note that any $\Ev({\xv})$ that satisfies Maxwell's equations is mode-volume-invariant under arbitrary scalar multiplication. Thus as a constraint we can instead impose the intensity at the origin to be 1, and leave the integral quantity in the numerator as the objective. This function is quadratic in the fields and linear in the permittivity, and leads to a tractable dual. The quadratic constraint on the intensity at the origin is itself cumbersome to work with in the duality formulation, and can be simplified by the observation that fixing a field intensity at a given point is equivalent to fixing its amplitude while allowing its direction and phase to vary freely. Thus, instead of imposing the quadratic constraint $\varepsilon(\xv_0)|\Ev(\xv_0)|^2=1$, we can instead set $\sqrt{\epso} \hat{\vect{p}} \cdot \Ev(\xv_0) = 1$, where $\vect{p}$ is a phase-dependent polarization vector, and $\epso$ is a binary value taking either of the two possible material-permittivity values. Thus we have ``lifted'' the problem to a higher dimension and converted the quadratic constraint to a linear one in the process. Therefore, we can rewrite \eqref{minproblem} as:
\begin{equation}
\begin{aligned}
& \underset{\varepsilon, \Ev, \hat{\vect{p}},\epso}{\text{minimize}}
& & \int \varepsilon |\Ev|^2 \\
& \text{subject to}
& & \nabla \times \nabla \times \Ev = \omega^2 \varepsilon \Ev.\\
& & & \sqrt{\epso} \hat{\vect{p}} \cdot \Ev(\xv_0) = 1
\end{aligned}
\label{eq:minproblem2}
\end{equation}

The next step is to discretize \eqref{minproblem2}. This can be done by any standard discretization scheme (finite elements, finite differences, etc.), and we use the finite-difference method~\cite{fdfdbook}. We add perfectly matched layers (PMLs) in the exterior region to truncate the computational domain. For a mode with high quality factor, $\Ev$ outside the resonator is very nearly zero (by definition), and thus we compute the integral in the mode volume calculation over the all space up to the PMLs. We write the curl-curl operator normalized by squared frequency, $(1/\omega^2) \nabla \times \nabla \times$, as $D$ in its discrete form. The number of degrees of freedom is $N\times M$, the total number of grid points, multiplied by 3 for the three electric-field components, such that $D \in \mathbb{C}^{3NM\times3NM}$. Let $e= \begin{pmatrix} e_x \\ e_y \\ e_z \end{pmatrix}$, $\varepsilon = \begin{pmatrix} \varepsilon_x \\ \varepsilon_y \\ \varepsilon_z \end{pmatrix}$ and ``diag'' represent the diagonal matrix with its argument on the diagonal and all zeros otherwise. We can then write the Maxwell constraint as,
\begin{equation}
\underbrace{\begin{bmatrix}
\Re(D) & -\Im(D) \\
\Im(D)  & \Re(D) 
\end{bmatrix}}_{\vect{A}}
\underbrace{\begin{pmatrix}
\Re(e) \\ \Im(e)
\end{pmatrix}}_e = 
\underbrace{\begin{bmatrix}
\diag{\varepsilon} & 0\\ 
0 & \diag{\varepsilon}
\end{bmatrix}}_{\diag{\varepsilon}}
\underbrace{\begin{pmatrix}
\Re(e) \\ \Im(e)
\end{pmatrix}}_e
\end{equation}
To properly compute $\int \varepsilon |\Ev|^2$ over the entire domain \emph{except} the PML region, we introduce a diagonal weight matrix $W$ that is unity everywhere except the PML, where it takes a small but nonzero value (important for invertibility below). Denoting the differential volume (area) of each grid cell as $S$, we can then write our objective function as 
\begin{equation}
\underbrace{\begin{pmatrix}
\Re(e) \\ \Im(e)
\end{pmatrix}^T}_{e}
\underbrace{\begin{bmatrix}
\diag{W} & 0 \\
0  & \diag{W}
\end{bmatrix}}_W
\underbrace{\begin{bmatrix}
\diag{\varepsilon} & 0\\ 
0 & \diag{\varepsilon}
\end{bmatrix}}_{\diag{\varepsilon}}
\underbrace{\begin{bmatrix}
\diag{W} & 0 \\
0  & \diag{W} 
\end{bmatrix}}_W
\underbrace{\begin{pmatrix}
\Re(e) \\ \Im(e)
\end{pmatrix}}_e\times S
\end{equation}
(Since $S$ is just a constant we do not include it in the optimization and only reintroduce in the final calculations.) Our optimization problem is now
\begin{equation}
\begin{aligned}
& \underset{\varepsilon, \epso, e, v}{\text{minimize}}
& & e^T W^T \diag{\varepsilon} W e \\
& \text{subject to}
& & A e  = \diag{\varepsilon} e \\
& & & \sqrt{\epso} v^T e = 1,
\end{aligned}
\label{eq:genminproblem}
\end{equation}
where $v$ is the representation of a delta function at $\xv_0$ with amplitude $\hat{\vect{p}}$ (separated into real and imaginary parts) in our finite-dimensional basis. For 2D TE mode ($\Ev$ out of plane), $v$ is just the projection vector that select the real part of $\Ev_z$, in which it is a constant. This is because for any TE mode, we can always pick a global phase such that $\Ev_z$ is real at origin. For 2D TM mode ($\Ev$ in-plane), $v$ may depends on two parameters, polarization ($\phi$) and the phase difference ($\delta$). Here, we pick the global phase such that $\Ev_x$ at origin is real.  We can then write $v$ as,
\begin{equation}
v(\phi,\delta) = 
\begin{pmatrix}
u\cos{\phi} \\ 0\\u \sin{\phi}\cos{\delta} \\ u \sin{\phi}\sin{\delta}
\end{pmatrix}
\end{equation}
where $u\in \mathbb{R}^{NM\times1}$ is the unit vector that select the origin (1 at the origin and 0 otherwise). Similarly, it extends to the three dimensional cases in which $v$ will depend on 3 parameters. Since $\epso{}$ and $v$ occupy a low-dimensional space, we can treat them as a parameters in the ``inner loop'' of the optimization where we optimize over $e$ and $\varepsilon$, and then optimized themselves in an ``outer loop'' search (See \Aref{scheme}).

Treating $v$ and $\epso{}$ as parameters and concatenating the two linear constraints together, we get the our final form of the primal problem. 
\begin{equation}
\begin{aligned}
& \underset{\varepsilon, e}{\text{minimize}}
& & e^T W^T \diag{\varepsilon} W e \\
& \text{subject to}
& & \underbrace{\begin{bmatrix} A \\ \sqrt{\varepsilon_{\rm origin}} v^T\end{bmatrix}}_A e - \begin{bmatrix} 
\diag{\varepsilon}\\ 0 \end{bmatrix} e= \underbrace{\begin{pmatrix}0 \\ 1\end{pmatrix}}_b
\end{aligned}
\label{eq:genminproblem}
\end{equation}
We then find the dual problem of \eqref{genminproblem} using standard procedures described in \cite{Boyd2004}. The Lagrangian of the problem is,
\begin{equation}
L(\varepsilon,e,\lambda) = e^T W^T \diag{\varepsilon} W e  + \lambda^T\left(\left(A-\begin{bmatrix}\diag{\varepsilon}\\0\end{bmatrix}\right)e-b\right)+I(\varepsilon)
\label{eq:lagrangian}
\end{equation}
where $\lambda$ is the dual variable, $I$ is an indicator function that is zero for valid permittivity distribution and $+\infty$ otherwise. The dual function g($\lambda$) is defined as the minimum of the Lagrangian $L(\varepsilon,e,\lambda)$ over variable $\varepsilon$ and $e$. To minimize $L$ over $e$, we can simply set $\frac{\partial{L}}{\partial{e}} = 0$:
\begin{equation}
\frac{\partial{L}}{\partial{e}} = 2W \diag{\varepsilon} W^Te + (A^T-\begin{bmatrix}\diag{\varepsilon}&0\end{bmatrix})\lambda) = 0,
\end{equation}
which implies that the optimal $e$, denoted $e^*$, is given by
\begin{equation}
    e^* = -\frac{1}{2}(W \diag{\varepsilon} W^T)^{-1}(A^T-\begin{bmatrix}\diag{\varepsilon}&0\end{bmatrix})\lambda.
\end{equation}
Hence,
\begin{equation}
\begin{aligned}
g(\lambda) &= \inf_{\varepsilon}\inf_{e}L(\varepsilon,e,\lambda)\\
           & = \inf_{\varepsilon} \left\{ -\frac{1}{4}\lambda^T\left(A-\begin{bmatrix}\diag{\varepsilon}\\0\end{bmatrix}\right)(W \diag{\varepsilon} W^T)^{-1}(A^T-\begin{bmatrix}\diag{\varepsilon}&0\end{bmatrix})\lambda - \lambda^Tb +I(\varepsilon)\right\}\\
           & = \inf_{\varepsilon^{(1)}\leq\varepsilon\leq\varepsilon^{(2)}}\left\{ -\frac{1}{4}\left\|W^{-1} \diag{\frac{1}{\sqrt{(\varepsilon)}}}(A^T-\begin{bmatrix}\diag{\varepsilon}&0\end{bmatrix})\lambda\right\|^2 - \lambda^Tb \right\} \\
           & = \inf_{\varepsilon^{(1)}\leq\varepsilon\leq\varepsilon^{(2)}} \left\{ -\frac{1}{4} \sum_i \sum_{j\in S_i} W_{j,j}^{-2}\frac{((A^T\lambda)_j-\varepsilon_i\lambda_j)^2}{\varepsilon_i} - \lambda^T b \right\} \\
& = -\frac{1}{4}\sum_i \max\left\{\sum_{j\in S_i}\frac{\left((A^T\lambda)_j-\varepsilon_i^{(2)}\lambda_j\right)^2}{W_{j,j}^2\varepsilon_i^{(2)}},\sum_{j\in S_i}\frac{\left((A^T\lambda)_j-\varepsilon_i^{(1)}\lambda_j\right)^2}{W_{j,j}^{2}\varepsilon_i^{(1)}}\right\} - \lambda^T b.
\end{aligned}
\label{eq:dualfunction}
\end{equation}
Note that to match the sizes of the other variables, the vector $\varepsilon$ has $6NM$ entries, but only $NM-1$ degrees of freedom, as we assume isotropic nonmagnetic media (reducing $6NM$ to $NM$), and the value at the origin is fixed to either $\epso^{(1)}$ or $\epso^{(2)}$. To ensure that the permittivities are constrained accordingly, we define sets $S_i$ that contains all elements in $\varepsilon$ that are equivalent to each other, independently denoted as $\varepsilon_i$. We first group all terms with free variable $\varepsilon_i$ together, i.e. sum over $S_i$. This separates each $\varepsilon_i$ and allows us to optimize each variable independently. During the derivation, we use the observation that the maximum of $\sum_i{(a_i+b_i\varepsilon)^2}/{\varepsilon}$ over $\varepsilon\in[\varepsilon^{(1)}, \varepsilon^{(2)}]$ always occur at the boundary---this is true as when $\varepsilon>0$, the second derivative of $\sum_i{(a_i+b_i\varepsilon)^2}/{\varepsilon}$ is $\sum_i{2a_i^2}/{\varepsilon^3}>0$. By introducing an auxiliary variable t, we can therefore write our dual problem, $\max_{\lambda} g(\lambda)$, as:
\begin{equation}
\begin{aligned}
& \underset{t, \lambda}{\text{maximize}}
& & -\frac{1}{4}\vect{1}^T t-\lambda^T b\\
& \text{subject to} 
& & t_i\geq\sum_{j\in S_i}\frac{((A^T\lambda)_j-\varepsilon_i^{(k)}\lambda_j)^2}{{W_{j,j}^2\varepsilon_i^{(k)}}}, \\ 
& & &\quad \forall i\in[NM], \quad k\in\{1,2\}.
\end{aligned}   
\label{eq:dual}
\end{equation}

We can implement the minimal feature size by grouping several discretization cells together to form a larger fabrication block. The derivation is exactly the same as the one shown in \eqref{dualfunction}. All $\varepsilon$ values within a fabrication block are set to be the same, i.e. the sets $S_i$ now additionally contain all elements in $\varepsilon$ that are in the same fabrication block $i$, and we can minimize over the fewer remaining degrees of freedom. Let $m$ be the total number of fabrication blocks. The dual problem that implement the minimal feature size is thus,
\begin{equation}
\begin{aligned}
& \underset{t, \lambda}{\text{maximize}}
& & -\frac{1}{4}\vect{1}^T t-\lambda^T b\\
& \text{subject to} 
& & t_i\geq\sum_{j\in S_i}\frac{((A^T\lambda)_j-\varepsilon_i^{(k)}\lambda_j)^2}{{W_{j,j}^2\varepsilon_i^{(k)}}}, \\ 
& & &\quad \forall i\in[m], \quad k\in\{1,2\}.
\end{aligned}   
\label{eq:dualblock}
\end{equation}
For very small mode volumes the bounds are dominated by the features closest to the origin, as expected. Therefore, for practical purposes, in our calculations we only implement minimal feature size at the center (only $\varepsilon$ within the central fabrication block need to be equal to each other).

Finally, \eqreftwo{dual}{dualblock} are convex QCQP problems that can be solved using standard convex optimization techniques~\cite{Boyd2004}. In our calculation, we first use CVX to rewrite it as a second-order cone problem, then we solve it using Gurobi solver~\cite{gurobi,cvx}. The whole optimization scheme to find out the minimal mode volume for TM mode is shown in in \Aref{scheme}. The analogous scheme also works for 3D cases. \\
\begin{algorithm}[H]
\SetKwFunction{SolveDual}{SolveDual}
\SetKwData{dualbound}{dualbound} \SetKwData{bound}{bound} \SetKwData{S}{S}
\bound = $+\infty$ \\
\For{$\epso \in[\varepsilon_1, \varepsilon_2]$}
    {\For{$\phi \in [0,\frac{\pi}{4}]$, $\delta \in [0, 2\pi]$}
        {\dualbound $\leftarrow$ \SolveDual($\epso$, $\phi$, $\delta$)\;
        \lIf{\dualbound $<$ \bound}{\bound = \dualbound}
        }
    }
\KwRet{\bound$\times$ \S}
\label{a:scheme}
\end{algorithm}

\begin{figure}[H]
    \centering
    \includegraphics[width=0.4\linewidth]{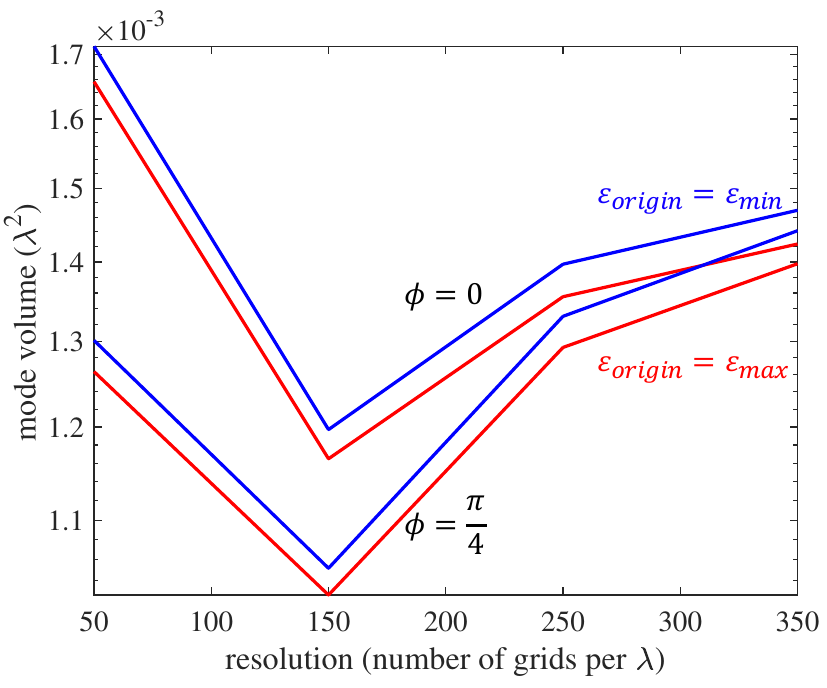}
    \caption{Bound calculated with different $\epso$ and different source parameters $(\phi,\delta)$ with same fabrication constraints for 2D TM mode. We observe that the bounds converge to similar values as resolution increases.}
    \label{fig:converge}
\end{figure}

As explained in algorithm \Aref{scheme}, we loop over $(\phi,\delta)$ to find the best (lowest) bound among all source polarization directions and phases. In fact, we find that this loop is not necessary for high enough resolution (and small minimum feature sizes). One can expect that if the fabrication constraints, domain, and PMLs all had circular symmetry, then the symmetry would \emph{require} equivalent bounds in each scenario. Our finite-difference discretization inhibits any such symmetrization, but for high enough resolution we nevertheless find a convergence of all of these bounds, as depicted in \figref{converge}. Intriguingly, the symmetry arguments do not necessarily imply that the bounds should be equivalent when the permittivity at the origin is switched between $\epso^{(1)}$ and $\epso^{(2)}$, and yet computational experiments show that to be true as well.
\section{Bound as a function of imag($\omega$)}
\begin{figure}[H]
    \centering
    \includegraphics[width=0.8\linewidth]{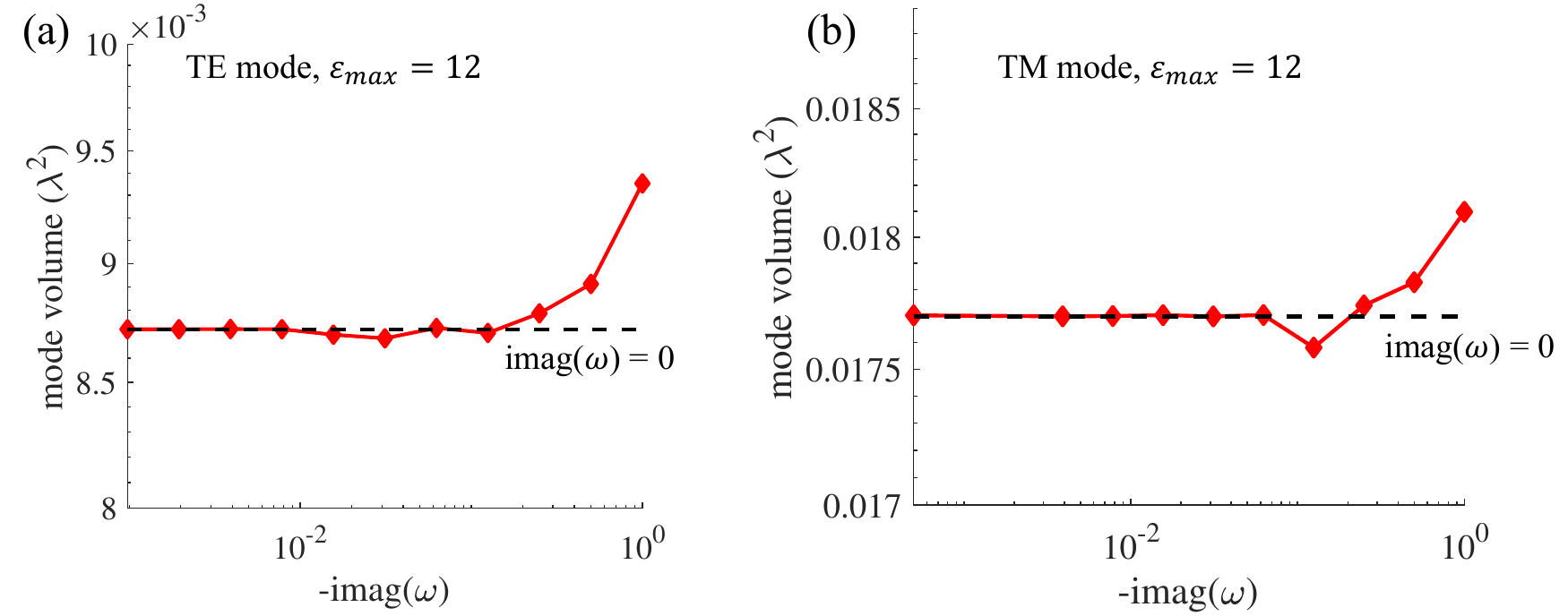}
    \caption{Lower bounds for mode volume as a function of imag($\omega$). For both (a) TE and (b) TM mode, the mode volume converges when imag($\omega$) goes to zero and converges to the bound value where imag($\omega$) is setted to be 0 (black dashed line).}
    \label{fig:imagw}
\end{figure}

An important technical point in our specification of the problem, \eqreftwo{minproblem}{genminproblem}, is that we specify the resonant frequency $\omega$ to be the real-valued frequency of interest, with no imaginary part. Any resonant cavity has a nonzero imaginary frequency, which would seem to invalidate such an approach. However, we compute bounds along a sequence of complex frequencies with imaginary parts tending to zero, and in numerical experiments they converge to their smallest values at 0, as shown in \figref{imagw}. The small dips below the asymptotic values appear to be numerical artifacts that do not persist as resolution increases.

\section{Bound as a function of design region}
The mode-volume bounds rapidly converge as a function of the diameter of the designable region beyond roughly a single wavelength. A similar observation is also seen in \citeasnoun{Liang2013}.
\begin{figure}[H]
    \centering
    \includegraphics[width=0.8\linewidth]{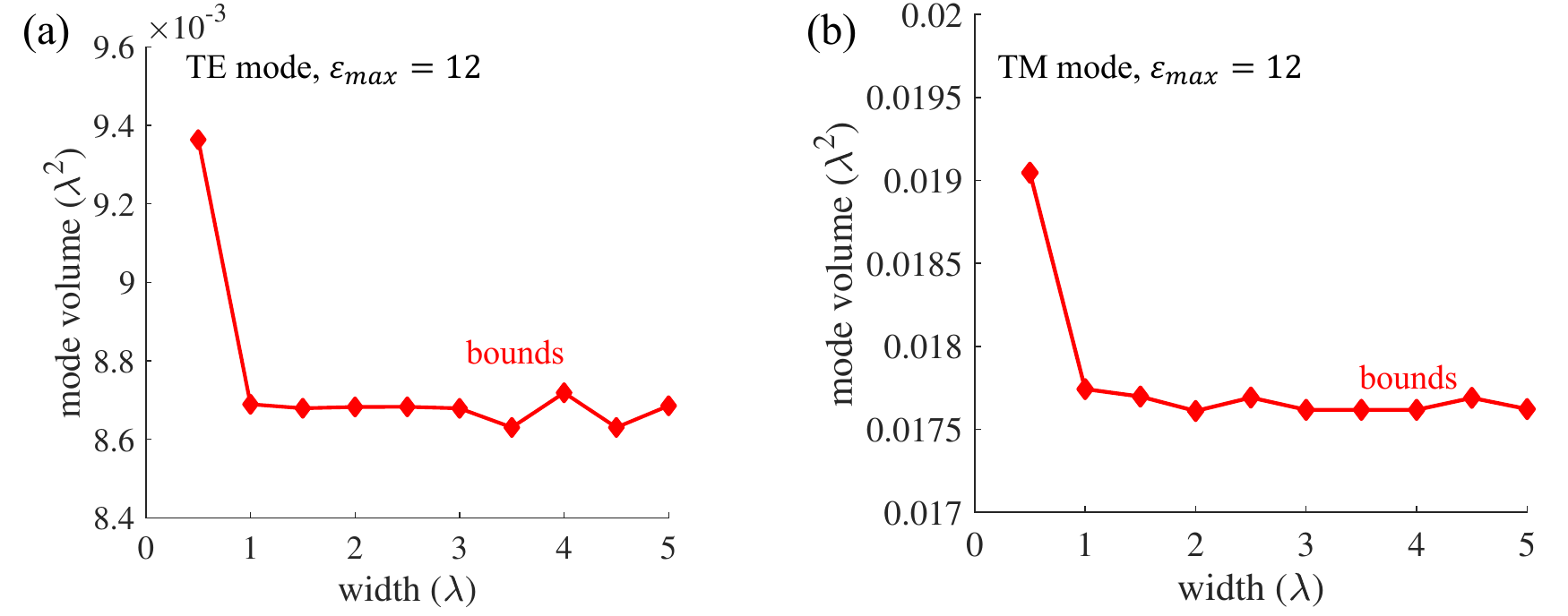}
    \caption{Lower bounds for mode volume as a function of diameter of the designable region. For both (a) TE and (b) TM mode, the mode volume converges rapidly when the designable region exceed $1\lambda \times 1\lambda$.}
    \label{fig:width}
\end{figure} 

\section{Average mode volume for multimode cavities}
The multimode cavity problem can be formulated by adding the integral-equation objectives while incorporating all constraints at each frequency at once. Some caution is needed in forming the objective. For TE modes, where the bounds typically scale as $\lambda^2$, simply summing the integrals at each individual frequency leads to a bias towards reducing only the longer-wavelength mode that dominates the objective. To circumvent this, we scale the terms in the objective by the squares of the frequency (for 2D problems) to counteract this bias. For example, our formulation of the two-frequency optimization problem is:
\begin{equation}
\begin{aligned}
& \underset{\varepsilon, \epso, e_1, e_2, v}{\text{minimize}}
& & \omega_1^2 (e_1^T W^T \diag{\varepsilon} W e_1) + \omega_2^2 (e_2^T W^T \diag{\varepsilon} W e_2) \\
& \text{subject to}
& & A_1 e_1  - \diag{\varepsilon} e_1 = 0\\
& & & A_2 e_2  - \diag{\varepsilon} e_2 = 0\\
& & & \sqrt{\epso} v^T e_1 = 1 \\
& & & \sqrt{\epso} v^T e_2 = 1,
\end{aligned}
\label{eq:average}
\end{equation}
By taking $\begin{pmatrix} A_1 & 0 \\ 0 & A_2 \end{pmatrix} \rightarrow A$, $\begin{pmatrix} e_1 \\ e_2 \end{pmatrix} \rightarrow e$, $\begin{pmatrix}{\omega_1}W & 0 \\ 0 & {\omega_2} W \end{pmatrix} \rightarrow W$, $\begin{pmatrix}\varepsilon \\ \varepsilon\end{pmatrix} \rightarrow \varepsilon$, we can rewrite \eqref{average} as \eqref{average2} which is of exact same form as \eqref{genminproblem}. Hence the corresponding dual problem can be derived using similar procedures described in Section I.
\begin{equation}
\begin{aligned}
& \underset{\varepsilon, e}{\text{minimize}}
& & e^T W^T \diag{\varepsilon} W e \\
& \text{subject to}
& & \begin{bmatrix} \multicolumn{2}{c}{\mathcal{A}} \\ \sqrt{\varepsilon_{\rm origin}} v^T & 0 \\ 0 & \sqrt{\varepsilon_{\rm origin}} v^T \end{bmatrix} e - \begin{bmatrix} \diag{\varepsilon}\\ 0 \\ 0\end{bmatrix} e= \begin{pmatrix}0 \\ 1 \\1 \end{pmatrix}
\end{aligned}
\label{eq:average2}
\end{equation}

\section{Inverse design of microcavities for minimal mode volumes}

\subsection{Optimization scheme}
\subsubsection{Converting non-Hermitian eigen-problems to scattering problems via LDOS}
Mathematically, non-Hermitian eigen-problems can be difficult to solve and sometimes even ill-defined. The resonances inside the open system under investigation are not normal modes and don't possess normalizable eigen-functions. Instead of solving for the quasi-normal modes, we identify local density of states (LDOS), defined relative to the total power $P$ generated by a point source at position $\bm{x}'$ polarized in direction $\ell$, as our figure of merit (FoM):
\begin{equation}\label{eq:ldos}
\begin{aligned}
\textrm{LDOS}_l(\omega,\bm{x}') &= \frac{4}{\pi}P \\
&= -\frac{2}{\pi} \Re \int \bm{J}^*(\bm{x})\cdot\bm{E}(\bm{x})d\bm{x}\\
&= -\frac{2\omega}{\pi}\Im G_{ll}(\bm{x}',\bm{x}'),
\end{aligned}
\end{equation}

where $\bm{G}_l(\bm{x},\bm{x}')$ is the Maxwell Green's function, satisfying
\begin{equation}
\begin{aligned}
(\nabla\times\frac{1}{\mu}\nabla\times-\omega^2\epsilon)\bm{G}_l(\bm{x},\bm{x}') = \hat{\bm{e}}_l \delta(\bm{x}-\bm{x}').
\end{aligned}
\end{equation}
The LDOS is proportional to the enhancement of the spontaneous-emission rate, and for a single-mode resonator (or a resonator with well-spaced modes), is proportional to the Purcell enhancement factor, and in particular the quantity $Q_i / V_i$, where $Q_i$ and $V_i$ are the quality factor and mode volume, respectively, of mode $i$:
\begin{equation}\label{eq:purcell}
\begin{aligned}
\textrm{LDOS}_{i} = \frac{2}{\pi \omega_i \epsilon(\bm{x}')}\frac{Q_i}{V_i}.
\end{aligned}
\end{equation}

Eq. \ref{eq:ldos} and Eq. \ref{eq:purcell} show how LDOS connects the possibly formidable non-Hermitian eigen-problem to a scattering problem where solving linear systems of equations would be sufficient. In this spirit, we maximize LDOS of the cavity in a finite system, and look for designs with minimal mode volumes by constraining quality factors.

\subsubsection{Limiting high quality factors by bandwidth averaging}
In order to minimize mode volume by maximizing LDOS, the optimization program should be constrained to designs with limited quality factor. Meanwhile, there is no use in practical cavities to increase the quality factors beyond certain point, which is determined by the loss budget of the system. 
Bandwidth-averaging is equivalent to bringing artificial material absorption to the system, which not only limits quality factors, but also automatically avoids poles in the Green's function. When the radiation loss is smaller than this material absorption, the overall quality factor $Q_{\rm total}$ has much room to improve if the quality factor from radiation $Q_{\rm rad}$ is improved. As $Q_{\rm rad}$ is large enough - comparable to $Q_{\rm abs}$, the optimizer would tend to minimize mode volume to further increase LDOS. Indeed we observe that in the first phase of the optimization, the mode volume V tends smaller and the quality factor Q tenders larger, before the decreasing of V becomes dominant as Q saturates the given artificial material quality factor. 

\subsection{Optimized designs}

\subsubsection{TM-mode cavities}
\begin{center}
\begin{tabular}{ | m{3.5cm} | m{3.5cm}| m{5cm} | }
\hline
   \hspace{3mm} feature size, d ($\lambda$)  & \hspace{2mm} mode volume ($\lambda^2$) &  \hspace{18mm} design\\
    \hline
    \hspace{12mm} 1/50 &   \hspace{12mm} 0.0056 & \vspace{1mm}\hspace{7mm} \phototm{TM150_grid3} \\
    \hspace{12mm} 3/100 &   \hspace{12mm} 0.0092 & \vspace{1mm}\hspace{6.5mm} \phototm{TM100_grid3} \\
    \hspace{12mm} 1/20 &   \hspace{12mm} 0.018 & \vspace{1mm}\hspace{6.5mm} \phototm{TM100_grid5} \\
    \hspace{12mm} 7/100 &   \hspace{12mm} 0.031 & \vspace{1mm}\hspace{6.5mm} \phototm{TM100_grid7} \\
    \hspace{12mm} 9/100 &   \hspace{12mm} 0.045 & \vspace{1mm}\hspace{6.5mm} \phototm{TM100_grid9} \\
    \hspace{12mm} 11/100 &   \hspace{12mm} 0.056 & \vspace{1mm}\hspace{6.5mm} \phototm{TM100_grid11} \\
\hline
\end{tabular}
\end{center}

\subsubsection{TE-mode cavities}
\begin{center}
\begin{tabular}{ | m{3.5cm} | m{3.5cm}| m{4cm} | }
\hline
   \hspace{3mm} feature size, d ($\lambda$)  & \hspace{2mm} mode volume ($\lambda^2$) &  \hspace{14mm} design\\
    \hline
    \hspace{12mm} 1/40 &   \hspace{12mm} 0.0077 & \vspace{1mm}\hspace{7mm} \photo{TE120_grid3} \\
    \hspace{12mm} 3/100 &   \hspace{12mm} 0.0077 & \vspace{1mm}\hspace{7mm} \photo{TE100_grid3} \\
    \hspace{12mm} 1/30 &   \hspace{12mm} 0.0077 & \vspace{1mm}\hspace{7mm} \photo{TE90_grid3} \\
    \hspace{12mm} 1/20 &   \hspace{12mm} 0.0077 & \vspace{1mm}\hspace{7mm} \photo{TE100_grid5} \\
    \hspace{12mm} 1/15 &   \hspace{12mm} 0.0099 & \vspace{1mm}\hspace{7mm} \photo{TE105_grid7} \\
    \hspace{12mm} 7/100 &   \hspace{12mm} 0.0096 & \vspace{1mm}\hspace{7mm} \photo{TE100_grid7} \\
    \hspace{12mm} 9/100 &   \hspace{12mm} 0.016 & \vspace{1mm}\hspace{7mm} \photo{TE100_grid9} \\
\hline
\end{tabular}
\end{center}

\bibliography{paper}